\documentclass[useAMS,usenatbib,usegraphicx]{mn2e}

\title[Two fossil groups of galaxies at $z \approx 0.4$ in the COSMOS]{Two fossil groups of galaxies at $z \approx 0.4$ in the COSMOS:\\ accelerated stellar-mass build-up, different progenitors}

\author[D. Pierini et al.]{D. Pierini$^{1}$\thanks{Currently guest astronomer at the MPE}, S. Giodini$^{2}$, A. Finoguenov$^{1}$, H. B\"{o}hringer$^{1}$, E. D'Onghia$^{3}$,
\newauthor G. W. Pratt$^{4}$, J. D\'emocl\`es$^{4}$, M. Pannella$^{4}$, S. Zibetti$^{5}$, F. G. Braglia$^{6}$,
\newauthor M. Verdugo$^{1}$, F. Ziparo$^{1}$, A. M. Koekemoer$^{7}$, M. Salvato$^{8,9,1}$
\newauthor and the COSMOS Collaboration\\
$^{1}$Max-Planck-Institut f\"ur extraterrestrische Physik, Giessenbachstrasse, D-85748, Garching bei M\"unchen, Germany\\
$^{2}$Leiden Observatory, P.O. Box 9513, NL-2300 RA  Leiden, The Netherlands \\
$^{3}$Harvard-Smithsonian Center for Astrophysics, 60 Garden Street, Cambridge, MA 02138, U.S.A.\\
$^{4}$Laboratoire AIM, DAPNIA/Service d'Astrophysique, CEA/DSM, CNRS, Universit\'e Paris Diderot, B\^at. 709, CEA-Saclay,\\ \hskip0.2cm F-91191 Gif-sur-Yvette Cedex, France\\
$^{5}$Dark Cosmology Centre, Niels Bohr Institute, University of Copenhagen, Juliane Maries Vej 30, 2100 Copenhagen \O, Denmark\\
$^{6}$Department of Physics and Astronomy, University of British Columbia, 6224 Agricultural Road, V6T 1Z1 Vancouver BC, Canada\\
$^{7}$Space Telescope Science Institute, 3700 San Martin Drive, Baltimore MD 21218, U.S.A.\\
$^{8}$Max-Planck-Institut f\"{u}r Plasmaphysik, Boltzmannstrasse 2, D-85748 Garching, Germany\\
$^{9}$Caltech, 1200 East California Blvd, PMA 249-17, Pasadena, CA 91125, USA
}
\begin{document}

\date{Accepted ... Received ...; in original form ...}

\pagerange{\pageref{firstpage}--\pageref{lastpage}} \pubyear{2011}

\maketitle

\label{firstpage}

\begin{abstract}
We report on two fossil groups of galaxies at redshifts $z = 0.425$ and $0.372$
discovered in the Cosmic Evolution Survey ({\it COSMOS\/}) area.
Selected as X-ray extended sources, they have total masses ($M_{200}$)
equal to $1.9~(\pm 0.41) \times 10^{13}$
and $9.5~(\pm 0.42) \times 10^{13}~\mathrm{M}_{\sun}$, respectively,
as obtained from a recent X-ray luminosity--mass scaling relation.
The lower mass system appears isolated,
whereas the other sits in a well-known large-scale structure (LSS)
populated by 27 other X-ray emitting groups.
The identification as fossil is based on the $i$-band photometry
of all the galaxies with a photometric redshift consistent with that
of the group at the $2 \sigma$ confidence level
and within a projected group-centric distance equal to $0.5 R_{200}$,
and $i_{\mathrm AB} \le 22.5~\mathrm{mag}$ limited spectroscopy.
Both fossil groups exhibit high stellar-to-total mass ratios
compared to all the X-ray selected groups of similar mass
at $0.3 \le z \le 0.5$ in the {\it COSMOS\/}.
At variance with the composite galaxy stellar mass functions (GSMFs)
of similarly massive systems, both fossil group GSMFs are dominated
by passively evolving galaxies
down to $M^{stars} \sim 10^{10}~\mathrm{M}_{\sun}$
(according to the galaxy broad-band spectral energy distributions).
The relative lack of star-forming galaxies
with $10^{10} \le M^{stars} \le 10^{11}~\mathrm{M}_{\sun}$
is confirmed by the galaxy distribution in the $b - r$ vs $i$
colour--magnitude diagram.
Hence, the two fossil groups appear as more mature
than the coeval, similarly massive groups.
Their overall star formation activity ended rapidly
after an accelerated build up of the total stellar mass;
no significant infall of galaxies
with $M^{stars} \ge 10^{10}~\mathrm{M}_{\sun}$
took place in the last 3 to 6 Gyr.
This similarity holds although the two fossil groups are embedded
in two very different density environments of the LSS,
which suggests that their galaxy populations were shaped by processes
that do not depend on the LSS.
However, their progenitors may do so.
We discuss why the late merging of a compact group
is favoured over the early assembly as a formation scenario
for the isolated, low-mass fossil group.
\end{abstract}


\section{Introduction}

In early numerical simulations, the merging of a compact galaxy group
was found to lead to the formation of a single elliptical galaxy
(hence a ``fossil group'') in a few billion years (Barnes 1989)\footnote{These numerical simulations adopted a 1:4 ratio of luminous to dark matter and assumed that individual galaxies had dark halos.}.
The final product retains an X-ray emitting halo of hot gas
(Ponman \& Bertram 1993).
The whole scenario was proved to be plausible
by the discovery of the fossil group
RX\,J1340.6$+$4018 (Ponman et al. 1994).

The standard observational definition of a fossil system
is based on the existence of extended X-ray emission
with a bolometric luminosity
$L_{\mathrm{X}} > 10^{42}~h_{50}^{-2}~\mathrm{erg~s}^{-1}$
and of an R-band magnitude gap $\mathrm{\Delta} m_{12} > 2$
between the brightest and second brightest members
within $0.5 R_{200}$\footnote{$R_{\mathrm{\Delta}}$ is the radius within which the total mass density of a group/cluster is equal to $\mathrm{\Delta}$ times the critical density of the Universe ($\rho_{\mathrm{c}}$). Correspondingly, $M_{\mathrm{\Delta}} = \mathrm{\Delta}~\rho_{\mathrm{c}} (z)~(4 \pi / 3)~R_{\mathrm{\Delta}}^3$ is the total mass enclosed by $R_{\mathrm{\Delta}}$.} (Jones, Ponman \& Forbes 2000).
In theoretical studies, the analogous definition of a fossil system
is less straightforward (see e.g., D'Onghia et al. 2005 for a discussion).

On the basis of the existing observations,
mostly limited to the local Universe, fossil groups are expected
to be a considerable population among the gravitationally bound systems
of galaxies (Vikhlinin et al. 1999; Romer et al. 2000; Jones et al. 2003;
La Barbera et al. 2009; Voevodkin et al. 2010).
For instance, their estimated spatial density is equal to
$2.83 \times 10^{-6}~h_{75}^{3}~\mathrm{Mpc}^{-3}$
for an X-ray bolometric luminosity
above $0.89 \times 10^{42}~h_{75}^{-2}~\mathrm{erg~s}^{-1}$
(La Barbera et al. 2009).
X-ray properties and dark-matter (DM) content of fossil groups
are comparable to those of bright groups and poor clusters of galaxies,
with total masses of $\sim 10^{13}$--$10^{14}~\mathrm{M}_{\sun}$,
whereas their brightest members are typically as luminous
as a cluster cD galaxy ($M_{\mathrm{R}} < -22.5 + 5~\mathrm{log}~h_{50}$).

In recent high-resolution hydrodynamical cosmological simulations
in a $\mathrm{\Lambda}$ cold dark matter ($\mathrm{\Lambda}$CDM) universe,
fossil groups have already assembled half of their final mass
at redshifts $z \ge 1$, and subsequently they typically grow by minor mergers
only, whereas non-fossil systems of similar total mass on average form later
(D'Onghia et al. 2005; Dariush et al. 2007).
The early assembly leaves sufficient time for galaxies
with a luminosity $L \ge L^{\star}$
(assuming a Schechter 1976 luminosity function)
to merge into the central galaxy by dynamical friction,
which yields the magnitude gap defining fossil groups.
This ``early formation'' scenario is alternative
to the compact group merging scenario (Barnes 1989).

However, no difference seems to exist
in the distribution of neighboring faint galaxy density excess,
distance from the red sequence\footnote{In a given colour-magnitude diagram, the red sequence is the locus populated by galaxies dominated by old, passively evolving, stellar populations (see Arimoto \& Yoshii 1987). This locus is particularly tight if the two broad-band filters that define the colour bracket the so-called $4000~\mathrm{\AA}$ break. Among coeval galaxies, the old, passively evolving ones exhibit the reddest colours, if dust attenuation is neglected (see Pierini et al. 2004 for a discussion).}, structural parameters, or stellar population properties between the brightest central galaxies (BCGs)
in 25 fossil groups at $z \le 0.1$ and a sample of bright field ellipticals
with a similar distribution in optical luminosity and X-ray properties
(La Barbera et al. 2009).
These authors argue that this evidence undermines the idea
that fossil systems represent the ``end product'' of a physical mechanism
that occurred at high redshift; it rather suggests that fossil groups
are the final stage of mass assembly in the Universe.

Whether either interpretation is true
or fossil groups correspond to a transient phase in the life of a galaxy group
(von Benda-Beckmann et al. 2008; Dariush et al. 2010; Cui et al. 2011),
whether they form a distinct family of groups
(D'Onghia \& Lake 2004; Khosroshahi, Ponman \& Jones 2007)
or not (Sales et al. 2007; Zibetti, Pierini \& Pratt 2009; Sun et al. 2009;
D\'emocl\`es et al. 2010; Voevodkin et al. 2010; Cui et al. 2011),
for several reasons it is important to extend the present knowledge
of fossil systems beyond the local Universe and towards lower total masses.

If undisturbed for a very long time, fossil groups may represent
the ultimate example of hydrostatic equilibrium in viriliazed systems
(Vikhlinin et al. 1999).
This conclusion is supported by a recent detailed study of two such systems
based on high-quality XMM-{\em Newton} spectro-imaging data
(D\'emocl\`es et al. 2010).

In addition, fossil groups could represent the least massive,
baryonically closed systems (Mathews et al. 2005),
their baryon mass fraction being equal to the cosmic mean value\footnote{The cosmic mean value of the baryon mass fraction can be obtained e.g. from the five years measurements of the Cosmic Microwave Background (CMB) performed with the {\em Wilkinson Microwave Anisotropy Probe} ({\it WMAP5\/}; Dunkley et al. 2009). According to this estimate, it amounts to $0.171 \pm 0.009$.}.
If so, they could form a reference sample
with a recent, quiescent merging activity with which one can investigate
the origin of the average baryon deficiency within $R_{500}$
($\approx 0.7 R_{200}$) found in X-ray emitting groups at $z \le 1$
(Giodini et al. 2009, hereafter referred to as G09;
see also Teyssier et al. 2011).
In particular, two pair-matched samples of fossil groups
with a range in total mass and redshift, one with radio-loud BCGs
and the other with radio-quiet BCGs, could be used to estimate
the ``instantaneous'' boost of the entropy in the central regions of the ICM
that is due to radio-galaxy feedback (cf. Giodini et al. 2010).

From simple considerations on the hierachical nature
of structure growth via merging (Milosavljevi\'c et al. 2006),
fossil systems are expected to be more frequent (by a factor of ten)
at the group mass-scale than at the cluster mass-scale
($\sim 10^{14}$--$10^{15}~\mathrm{M}_{\sun}$).
This is confirmed by the most recent study of fossil groups
based on hydrodynamical resimulations of five regions
in the Millennium Simulation and on galaxy formation models
(Cui et al. 2011 and references therein).
Therefore fossil groups may also help in understanding the properties
of more massive bound systems at $z = 0$, especially if it turns out
that they foster the formation of the most massive progenitor of the BCG
of a cluster, as suggested by J\,0454$-$0309 at $z = 0.26$
(Schirmer et al. 2010; cf. Barnes 1989 and references therein).

During our ongoing studies of the X-ray selected groups
in the $2~\mathrm{deg}^2$ area of the Cosmic Evolution Survey
({\it COSMOS\/}; Scoville et al. 2007),
we have discovered two fossil groups at $z = 0.372$ and $z = 0.425$.
As demonstrated by G09 and Giodini et al. (2010),
the multi-wavelength data set of the {\it COSMOS\/}
offers a thus far unique opportunity to characterize X-ray selected groups
of galaxies up to $z \sim 1$.
This enables us not only to fully describe the galaxy populations
of the two fossil groups, but also to characterize their surrounding
large-scale structures (LSSs), and to speculate on their progenitors.
In Sect.~2 we describe selection and properties of these two groups.
Their galaxy stellar mass functions and colour--magnitude diagrams
are determined in Sect.~3.
Discussion of results and inference on evolutionary scenarios
appear in Sect.~4; conclusions are drawn in Sect.~5.

As in G09, we adopt a $\mathrm{\Lambda}$CDM cosmological model
($\mathrm{\Omega_m} = 0.258$, $\mathrm{\Omega_\Lambda} = 0.742$)
with $\mathrm{H}_0 = 72~\mathrm{km~s}^{-1}~\mathrm{Mpc}^{-1}$,
consistently with results from {\it WMAP5\/}
(Dunkley et al. 2009; Komatsu et al. 2009).
Thus, a redshift of 0.372 (0.425) corresponds to
a lookback time of $3.981~\mathrm{Gyr}$ ($4.409~\mathrm{Gyr}$),
a luminosity distance of $1961.1~\mathrm{Mpc}$ ($2298.2~\mathrm{Mpc}$),
an angular diameter distance of $1041.8~\mathrm{Mpc}$ ($1131.8~\mathrm{Mpc}$),
and a scale of $5.051~\mathrm{kpc}/^{\prime \prime}$
($5.487~\mathrm{kpc}/^{\prime \prime}$).

Hereafter magnitudes are expressed in the AB system
unless otherwise noted.

\section[]{Sample selection and properties}

\begin{figure*}
\vskip -0.35truecm
\includegraphics[width=120mm]{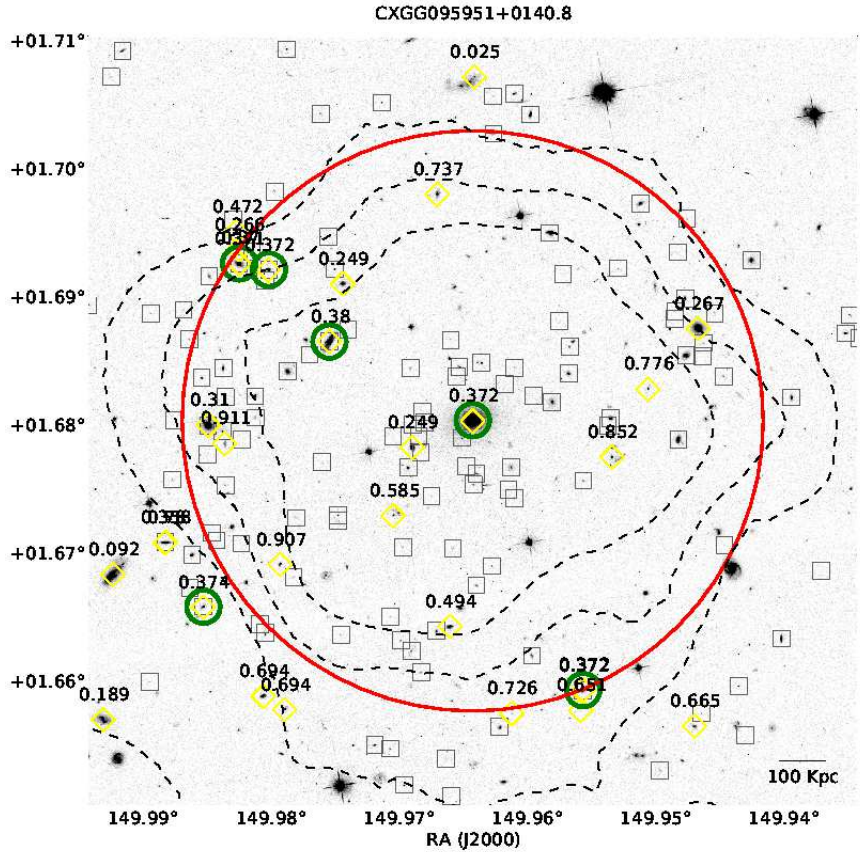}
\vskip -1.25truecm
\caption{$i$-band image of CXGG\,095951$+$0140.8, enclosing a circular area of radius $0.5 R_{200}$ centered on the X-ray position of the group (marked in red). Photometric members are marked with squares; green circles mark the spectroscopic members. Available spectroscopic redshifts of all objects in the region are given (yellow diamonds). Dashed contours indicate values of the X-ray surface brightness at the $4 \sigma$, $7 \sigma$ and $10 \sigma$ levels after removal of point-like sources, where $1 \sigma$ corresponds to $4.6 \times 10^{-16}~\mathrm{erg~s^{-1}~cm^{-2}~arcmin^{-2}}$ in the 0.5--$2~\mathrm{keV}$ band.}
\end{figure*}

\begin{figure*}
\vskip -0.35truecm
\includegraphics[width=120mm]{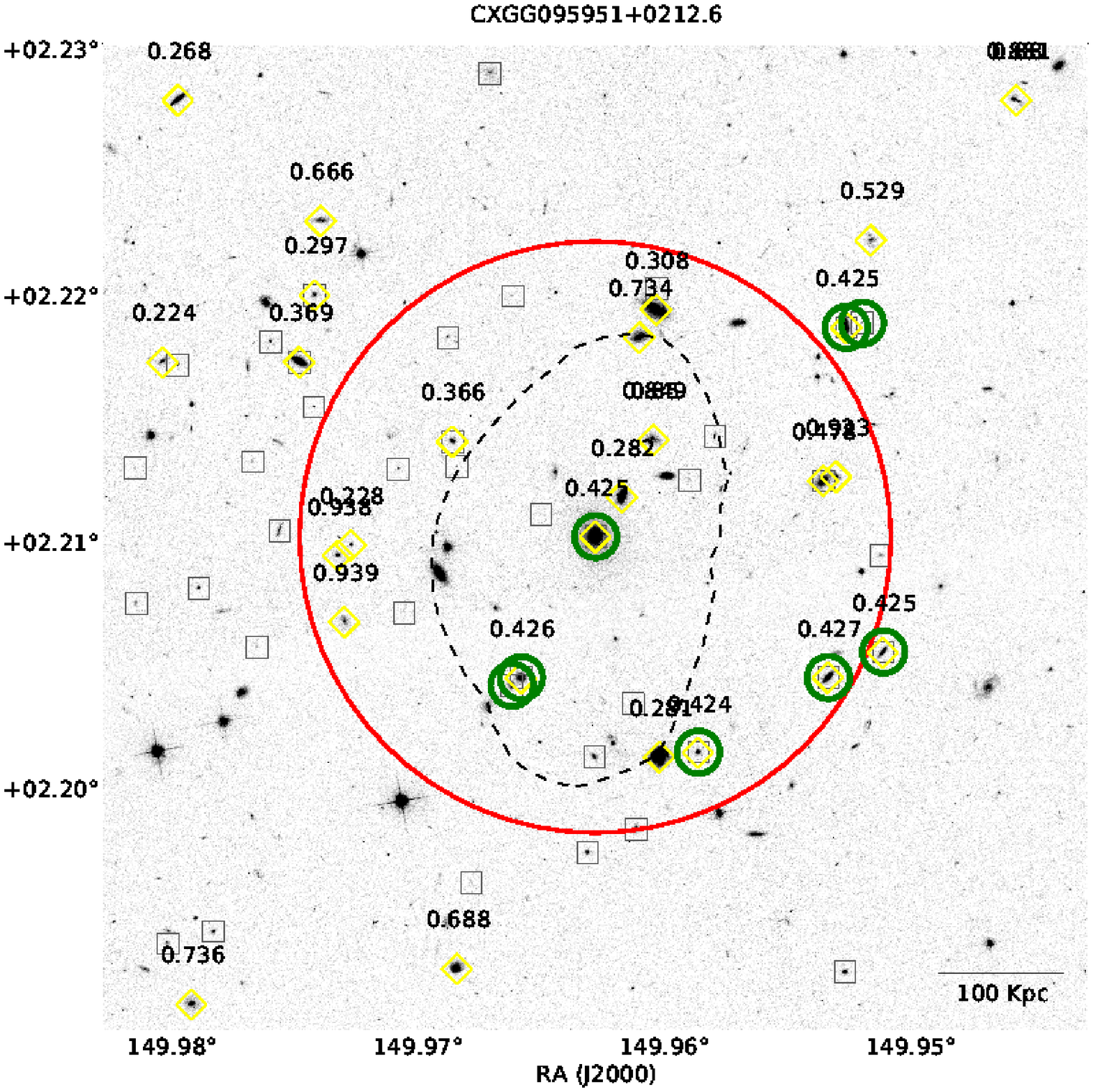}
\vskip -1.25truecm
\caption{The same as in Fig.~1 but for CXGG\,095951$+$0212.6. Here the $1 \sigma$ value of the X-ray surface brightness corresponds to $3.5 \times 10^{-16}~\mathrm{erg~s^{-1}~cm^{-2}~arcmin^{-2}}$ in the 0.5--$2~\mathrm{keV}$ band.}
\end{figure*}

From the composite mosaic of the XMM-{\em Newton} and {\em Chandra} X-ray data
mapping the {\it COSMOS\/} $2~\mathrm{deg}^2$ area,
extended sources (i.e., groups and clusters) were detected
down to a flux limit of $10^{-15}~\mathrm{erg~s}^{-1}~\mathrm{cm}^{-2}$
(Finoguenov et al. in preparation; see also Finoguenov et al. 2007).
G09 selected 91 groups/poor clusters at $0.1 < z < 1$,
which have an X-ray flux estimate at a significance better
than $3 \mathrm{\sigma}$, after point-like source subtraction
(see Finoguenov et al. 2007).
For these groups, the membership of a galaxy
(through a precise photometric redshift\footnote{For galaxies with $i \le 25~\mathrm{mag}$ at the redshift of the two fossil groups under study (i.e., $z \approx 0.4$), the $1 \sigma$ error on the photometric redshift is equal to $0.03 \times (1+z) = 0.042$.}) and the total stellar mass in galaxies
together with its uncertainty were robustly determined as described in G09.
The total mass of each group was determined
from the $L_{\mathrm{X}}$--$M_{200}$ relation
established through weak lensing analysis by Leauthaud et al. (2010).
Uncertainties on $M_{200}$ are statistical only
(i.e., they arise from the $L_{\mathrm{X}}$--$M_{200}$ relation).
The effects of systematics are estimated to be lower
than the statistical uncertainty\footnote{Systematic effects tested were: errors on photometric redshifts, mis-centering, uncertainties in the mass--concentration relation. These amount to a few percents (see fig.~7 in Leauthaud et al. 2010), while the statistical error is much larger. Removal of point-like sources can lead to underestimate the X-ray flux, since the scale at which they are removed matches with the scale of cool-cores. The correction, as detailed in sect.~4.2 of Leauthaud et al. (2010), amounts to 3--17\% of the total X-ray flux. This correction was taken into account when establishing the $L_{\mathrm{X}}$--$M_{200}$ relation.}.

We searched for fossil groups in the G09 sample.
As in Giodini et al. (in preparation), we defined as member candidates
of a group those galaxies within a projected group-centric distance
of $R_{200}$, and with a photometric redshift consistent with
the photometric redshift of the group within two times the uncertainty
on the photometric redshift of the group.
Then we inspected the brightest member-candidate galaxies
within a projected group-centric distance equal to $0.7 R_{200}$
to minimize the potential bias produced by uncertainties on $R_{200}$.
As for ranking in luminosity, the available $i$-band photometry was considered,
since the reference {\it COSMOS\/} photometric catalogue
is extracted from imaging in this band (Ilbert et al. 2009).
This choice avoids the use of template- and (photometric) redshift-dependent
$k$-corrections as for the choice of considering rest-frame R-band magnitudes.
Furthermore, up to $z = 1$, the $i$-filter (observed frame) maps emission
at rest-frame wavelengths redder than the 4000~\AA-break,
which limits the impact of young stellar populations
and, potentially, attenuation by a dusty interstellar medium.

A number of groups were initially selected as potential fossil systems,
the magnitude difference between the BCG
and any other bright member-candidate galaxy within $0.7 R_{200}$
being conservatively chosen to be larger than $1.5~i$-mag.
We set a larger search radius and a smaller magnitude gap
than those required by the definition of fossil groups (cf. Sect. 1)
in order to maximize the number of candidates
in this first round of the search for fossil groups.
Two groups were successively established as fossil
thanks to the existing spectroscopic information (Lilly et al. 2007),
limited to sources brighter than $i = 22.5~\mathrm{mag}$,
which enabled a finer screening against projected foreground/background
bright galaxies.

On one hand, we flagged out galaxies classified as member candidates
if they were located within $R_{200}$ from the X-ray centroid
of either fossil group but exhibited a line-of-sight velocity
in excess of $3000~\mathrm{km~s}^{-1}$ plus its $3 \sigma$ uncertainty\footnote{The uncertainty of the high-quality (flags 3 and 4) zCOSMOS spectroscopic redshifts amounts to $110~\mathrm{km~s}^{-1}$.} with respect to the spectroscopically determined recession velocity of either group.
The redshift of each group was computed
using the median spectroscopic redshifts of the galaxies within $R_{200}$
through iterative removal of outliers with a recession velocity
beyond $1500~\mathrm{km~s}^{-1}$ with respect to the recession velocity
of the group.
The starting redshift was defined
by the photometric redshifts of galaxies on the red sequence. 
Contamination was checked by running a red-sequence-finding algorithm
between redshifts 0 and 2 and by using spectroscopic group catalogs
(Knobel et al. 2009).
On the other hand, the available spectroscopy enabled us to check,
in particular, that bright galaxies within $0.5 R_{200}$
from the X-ray centroid of either fossil group were actually not dropped
as member candidates because of an inaccurate photometric redshift.
Additional considerations on size and luminosity were used to check
the goodness of the photometric redshifts of relatively bright galaxies
without spectroscopy (see CXGG\,095951$+$0212.6 in Fig.~2).

Figures 1 and 2 show cut-outs of the central regions
of the two {\it COSMOS\/} fossil groups under study,
which encompass the area within $0.5 R_{200}$ for each system.
They reproduce X-ray surface brightness contours
overlaid on the $i$-band HST/ACS images (Koekemoer et al. 2007)
of the two galaxy systems.
In particular, member candidates (i.e., photometric members) are indicated
together with all galaxies brighter than $i = 22.5~\mathrm{mag}$
having a spectroscopic redshift.
The BCG of either group can be easily identified
at the center of the X-ray emission region.

One system, CXGG\,095951$+$0140.8,
has $\mathrm{\Delta m_i}_{12} = 2.19 \pm 0.014~\mathrm{mag}$
(i.e., $\mathrm{\Delta m}_{12} = 2.10 \pm 0.02~\mathrm{mag}$
as obtained from the difference between the magnitudes
in the rest-frame $r$-band), $R_{200} = 832 \pm 19.6~\mathrm{kpc}$,
$M_{200} = 9.5~(\pm 0.42) \times 10^{13}~\mathrm{M}_{\sun}$,
$z = 0.372$, and six spectroscopic members.
It is part of a large-scale structure (see fig.~1 in G09
and fig.~3 in Scoville et al. 2007) populated by 28 X-ray emitting groups
distributed across the entire $2~\mathrm{deg}^2$ area of the {\it COSMOS\/}
(corresponding to a cross size of about $25.5~\mathrm{Mpc}$ at $z=0.37$).
Furthermore, its dominant galaxy (with a rest-frame absolute magnitude
$M_{i} = -24.87$) hosts a point-like radio source
(see Giodini et al. 2010).
The other fossil group, CXGG\,095951$+$0212.6,
has $\mathrm{\Delta m_i}_{12} = 2.35 \pm 0.014~\mathrm{mag}$
(i.e., $\mathrm{\Delta m}_{12} = 2.32 \pm 0.02~\mathrm{mag}$
in the rest-frame $r$-band), $R_{200} = 478 \pm 54.4~\mathrm{kpc}$,
$M_{200} = 1.9~(\pm 0.41) \times 10^{13}~\mathrm{M}_{\sun}$,
$z = 0.425$, and eight spectroscopic members.
It is isolated and its BCG has $M_{i} = -23.87$.

Basic properties of the two fossil groups under study are listed in Table 1.
In addition, we note that they populate the upper half of the distribution
of the {\it COSMOS\/} X-ray selected groups
in the (galaxy) stellar mass fraction--group total-mass diagram
(G09, their fig.~5), where quantities are estimated at $0.7 R_{200}$.
This is particularly true for CXGG\,095951$+$0212.6.
As from G09\footnote{Giodini et al. (in preparation) use a new catalogue of galaxy stellar masses by O. Ilbert instead of the catalogue by Ilbert et al. (2009) used in G09. The difference between the catalogues rests on the way stellar masses are computed. In the new one, stellar mass estimates are based on fitting the spectral energy distribution (SED) of a galaxy and not on the rest-frame K-band luminosity of the latter; furthermore, the stellar initial mass function of Chabrier (2003) is used instead of a Salpeter (1955) one. For the purposes of our study it is important that no systematics differently affects such estimates in fossil and non-fossil groups, not that the absolute scale of the estimated stellar mass of a galaxy is correct.}, the stellar mass fraction is equal to 0.039 for the more massive fossil group (CXGG\,095951$+$0140.8)
and 0.143 for the less massive one (CXGG\,095951$+$0212.6).
Thus, the fraction of the stellar mass in galaxies
decreases with increasing total mass
for the two fossil groups as for all other coeval groups (G09).

Furthermore, the more (less) massive fossil group exhibits a value
of the total $i$-band-to-X-ray luminosity ratio at $R_{200}$
($L_i/L_{{\mathrm X}}$) equal to 138 (305), after background subtraction
including spectroscopic information (from Giodini et al. in preparation).
Therefore CXGG\,095951$+$0140.8 (CXGG\,095951$+$0212.6) exhibits
a low (normal) value for its total mass with respect to the distribution
of $L_i/L_{{\mathrm X}}$ as a function of $M_{200}$ for the 51 groups
at $z=0.3$--0.5 considered by Giodini et al.\footnote{These COSMOS X-ray selected groups define a mass-complete sample, probing different large-scale density environments.}.

With a total mass of $1.9~(\pm 0.41) \times 10^{13}~\mathrm{M}_{\sun}$,
CXGG\,095951$+$0212.6 is among the least massive fossil systems
selected in X-rays, whereas CXGG\,095951$+$0212.6 is at the boundary between
the group and cluster mass regimes (cf. Vikhlinin et al. 1999;
Sun et al. 2009; Zibetti et al. 2009; D\'emocl\`es et al. 2010;
Voevodkin et al. 2010).
As a reference, RX\,J105453.3$+$552102 at $z = 0.47$, with a total mass of $\sim 10^{15}~\mathrm{M}_{\sun}$ (Aguerri et al. 2011), is among the most massive fossil systems\footnote{RX\,J105453.3$+$552102 is not considered as a fossil system by Aguerri et al. (2011), since its magnitude gap is equal to $1.92 \pm 0.09~\mathrm{mag}$ in the observed-frame $r$-band. The bright end of the galaxy luminosity function of this cluster is populated by old, passively evolving objects (Aguerri et al. 2011), so that the introduction of a $k$-correction is expected not to significantly modify the magnitude gap in the rest-frame $r$-band. However, the fossil criteria fixed by Jones et al. (2003) are somewhat arbitrary and depend on the way magnitudes are measured (see Aguerri et al. 2011 for a discussion). In addition, the likely contribution from diffuse light to the BCG emission is hard to be fully accounted for (see Zibetti et al. 2009 for a discussion). Thus, we consider RX\,J105453.3$+$552102 as a fossil system as we did for RXC\,J2315.7$-$0222 at $z = 0.026$ with $\mathrm{\Delta m}_{12} = 1.88 \pm 0.03~\mathrm{mag}$ in the rest-frame $\mathrm{R_C}$-band (D\'emocl\`es et al. 2010).}.

\begin{table*}
 \centering
 \begin{minipage}{150mm}
  \caption{Basic properties of the sample}
  \begin{tabular}{@{}lcccccrrc@{}}
  \hline
   Denomination & z & $\mathrm{M}_{200}$ & $\mathrm{R}_{200}$ & $\mathrm{\Delta m_i}_{12}$ & $\mathrm{\Delta m}_{12}$ & N & $\mathrm{N}_\mathrm{phot}$ & $\mathrm{N}_\mathrm{spec}$ \\
     & & [$10^{13}~\mathrm{M}_{\sun}$] & [kpc] & [mag] & [mag] \\
  \hline
 CXGG\,095951$+$0140.8 & 0.372 & 9.5 $\pm$ 0.42 & 832 $\pm$ 19.6 & 2.19 $\pm$ 0.014 & 2.10 $\pm$ 0.02 & 261 & 140 & 6 \\
 CXGG\,095951$+$0212.6 & 0.425 & 1.9 $\pm$ 0.41 & 478 $\pm$ 54.4 & 2.35 $\pm$ 0.014 & 2.32 $\pm$ 0.02 & 68 & 29 & 8 \\
  \hline
  \end{tabular}

  \medskip
  Denomination, spectroscopic redshift ($z$), total mass ($M_{200}$), size ($R_{200}$), observed $i$-band and rest-frame $r$-band magnitude gaps between the first two brightest member galaxies ($\mathrm{\Delta m_i}_{12}$ and $\mathrm{\Delta m}_{12}$, respectively), total number of galaxies  brighter than the completeness magnitude ($i = 25.37~\mathrm{mag}$ and $i = 25.17~\mathrm{mag}$ for CXGG\,095951$+$0140.8 and CXGG\,095951$+$0212.6, respectively) within $R_{200}$ ($N$), numbers of expected (i.e., the photometric members after statistical subtraction of field galaxies) and spectroscopic member galaxies within $R_{200}$ ($N_\mathrm{phot}$ and $N_\mathrm{spec}$, respectively) are listed for each fossil group.
 \end{minipage}
\end{table*}

\section[]{Results}

\subsection{Galaxy stellar mass function}

We compute the galaxy stellar mass function (GSMF) of the two fossil groups
in bins of 0.5 dex in stellar mass ($M^{\mathrm{stars}}$).
Only member-candidate galaxies with a stellar mass
above the completeness mass at the redshift of each group are considered.
The completeness mass is equal to $8.3 \times 10^8~\mathrm{M_{\sun}}$
or $1.25 \times 10^9~\mathrm{M_{\sun}}$ for, respectively,
all member-candidate galaxies of CXGG\,095951$+$0212.6
or those among them that are classified as passively evolving\footnote{{\it COSMOS\/} galaxies are classified as passively evolving or star forming on the basis of the template best-fitting their individual SEDs. In particular, the former galaxies have star formation rates lower than $10^{-2}~\mathrm{M_{\sun}~yr^{-1}}$.}.
It is equal to $3 \times 10^8~\mathrm{M_{\sun}}$
or $5 \times 10^8~\mathrm{M_{\sun}}$ for, respectively,
all member-candidate galaxies of CXGG\,095951$+$0140.8
or the passively evolving ones among them.

We applied a correction for the expected number of interlopers,
based on the statistics of the galaxies outside X-ray emitting
bound structures in the same photometric-redshift slice,
which define the coeval field of either group.
This holds whether or not galaxies are additionally selected
according to their star formation activity.
In each case, the field GSMF was normalized by the group-to-survey volume ratio
and then subtracted to the GSMF of the member-candidate galaxies
to obtain the GSMF of the (bona fide) member galaxies of each fossil group.
Counts per stellar mass bin are normalized to the total mass $M_{200}$
of each fossil group, to enable comparison with the composite GSMF
of the 40 (11) {\it COSMOS\/} X-ray selected groups at $0.3 \le z \le 0.5$
with $2 \times 10^{13} \le M_{200} \le 5 \times 10^{13}~\mathrm{M}_{\sun}$
($M_{200} > 5 \times 10^{13}~\mathrm{M}_{\sun}$) from Giodini et al.
(in preparation).
These two composite GSMFs are computed in an analogous way,
but include the number counts from all selected groups
with similar masses to that of either fossil group;
they are normalized to the sum of the total masses of the individual groups.

Figure 3 shows the normalized GSMFs of the two fossil groups
and the normalized composite GSMFs of the two reference samples
of X-ray selected groups at $0.3 \le z \le 0.5$ for all (bona fide)
member galaxies and for those classified as passively evolving.
Comparison of these normalized GSMFs reveals that about 80\% of the galaxies
with $M^{\mathrm{stars}} \ge 10^{10}~\mathrm{M}_{\sun}$ are passively evolving
in the two fossil groups at $z \approx 0.4$.
Conversely, passive galaxies start dominate the GSMF
of the coeval, mass-complete sample of X-ray selected groups
at much higher stellar masses, with a fraction of 80\% passive galaxies
being reached only at $M^{\mathrm{stars}} \ge 10^{11}~\mathrm{M}_{\sun}$.
This holds although the fraction of passively evolving galaxies
increases towards larger stellar masses across the entire range
in stellar mass.

Interestingly, half of the galaxies
with $M^{\mathrm{stars}} \sim 10^{11}~\mathrm{M}_{\sun}$ do not show evidence
of ongoing star-formation activity for similarly massive,
optically selected groups at $0.3 \le z \le 0.48$ (Wilman et al. 2008)\footnote{In Wilman et al. (2008), passive galaxies are defined on the basis of their {\em Spitzer}-IRAC mid-infrared colours, stellar masses are derived from rest-frame K-band luminosities (in analogy with Ilbert et al. 2009).}.

If the fraction of star-forming (``blue'') galaxies
is a proxy for the dynamical youth of a galaxy system at fixed total mass
(e.g., Tully \& Shaya 1984; Balogh, Navarro, \& Morris 2000;
Ellingson et al. 2001; Zapata et al. 2009; Braglia et al. 2009),
the previous difference suggests that
the two fossil groups assembled their DM haloes
earlier than the average system
among the similarly massive groups at $0.3 \le z \le 0.5$.

\begin{figure*}
\includegraphics[width=140mm]{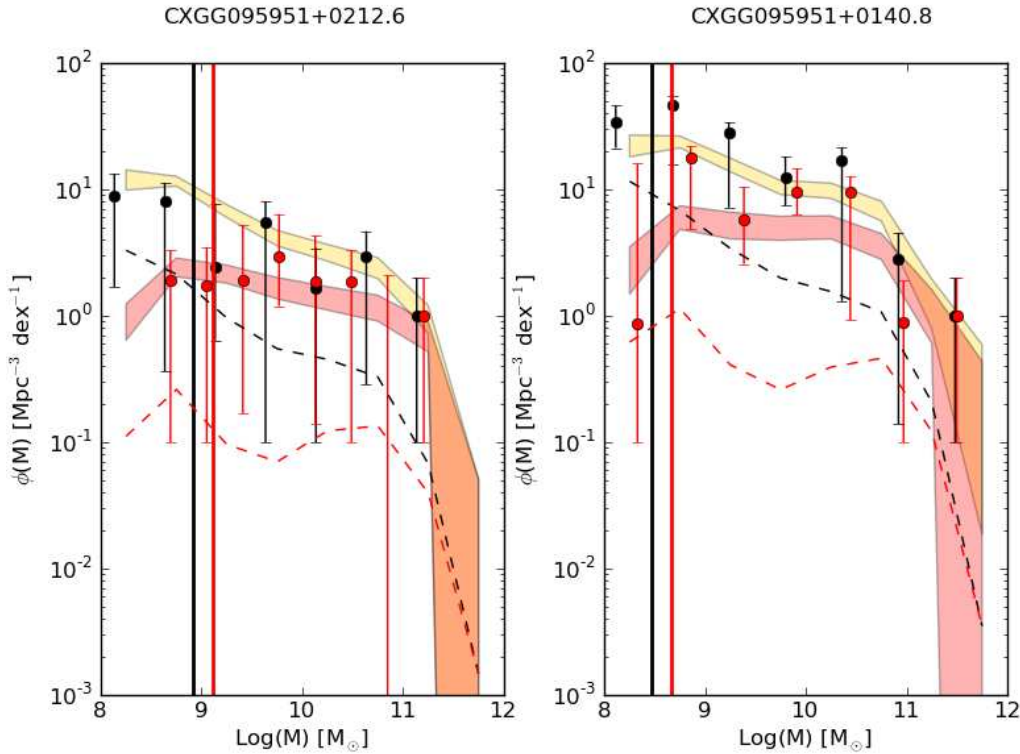}
\caption{The galaxy stellar mass functions (GSMFs) of the two fossil groups
CXGG\,095951$+$0212.6 (left) and CXGG\,095951$+$0140.8 (right).
Each panel shows the GSMF for all the (bona fide) member galaxies
(black circles) and the GSMF for the subsample of passively evolving
(bona fide) member galaxies (red symbols),
together with their statistical uncertainties.
Conversely, the vertical black and red lines represent, respectively,
the mass completeness limit for all the galaxies at the same redshift
as the group and for the passively evolving ones among them.
In each panel the black and red dashed curves represent, respectively,
the GSMFs for all the galaxies in the coeval field
and for the passively evolving ones among them.
To obtain the GSMF of all the (bona fide) member galaxies of either group,
the corresponding field GSMF was normalized by the group-to-survey volume ratio
and then subtracted to the GSMF of all the candidate members.
The GSMF of the passively evolving (bona fide) member galaxies of either group
was obtained in an analogous way.
Furthermore, each panel shows the composite GSMFs
of the coeval, X-ray selected groups for all their (bona fide) member galaxies
(yellow area) and the passively evolving ones among them (red area)
together with their statistical uncertainties (Giodini et al. in preparation).
Among the X-ray selected groups at $0.3 \le z \le 0.5$ in {\it COSMOS\/},
there are 40 systems as massive as CXGG\,095951$+$0212.6 (i.e., with $2 \times 10^{13} \le M_{200} \le 5 \times 10^{13}~\mathrm{M}_{\sun}$) and 11 systems
as massive as CXGG\,095951$+$0140.8 (i.e., with $M_{200} > 5 \times 10^{13}~\mathrm{M}_{\sun}$).}
\end{figure*}

\begin{figure*}
\includegraphics[width=100mm]{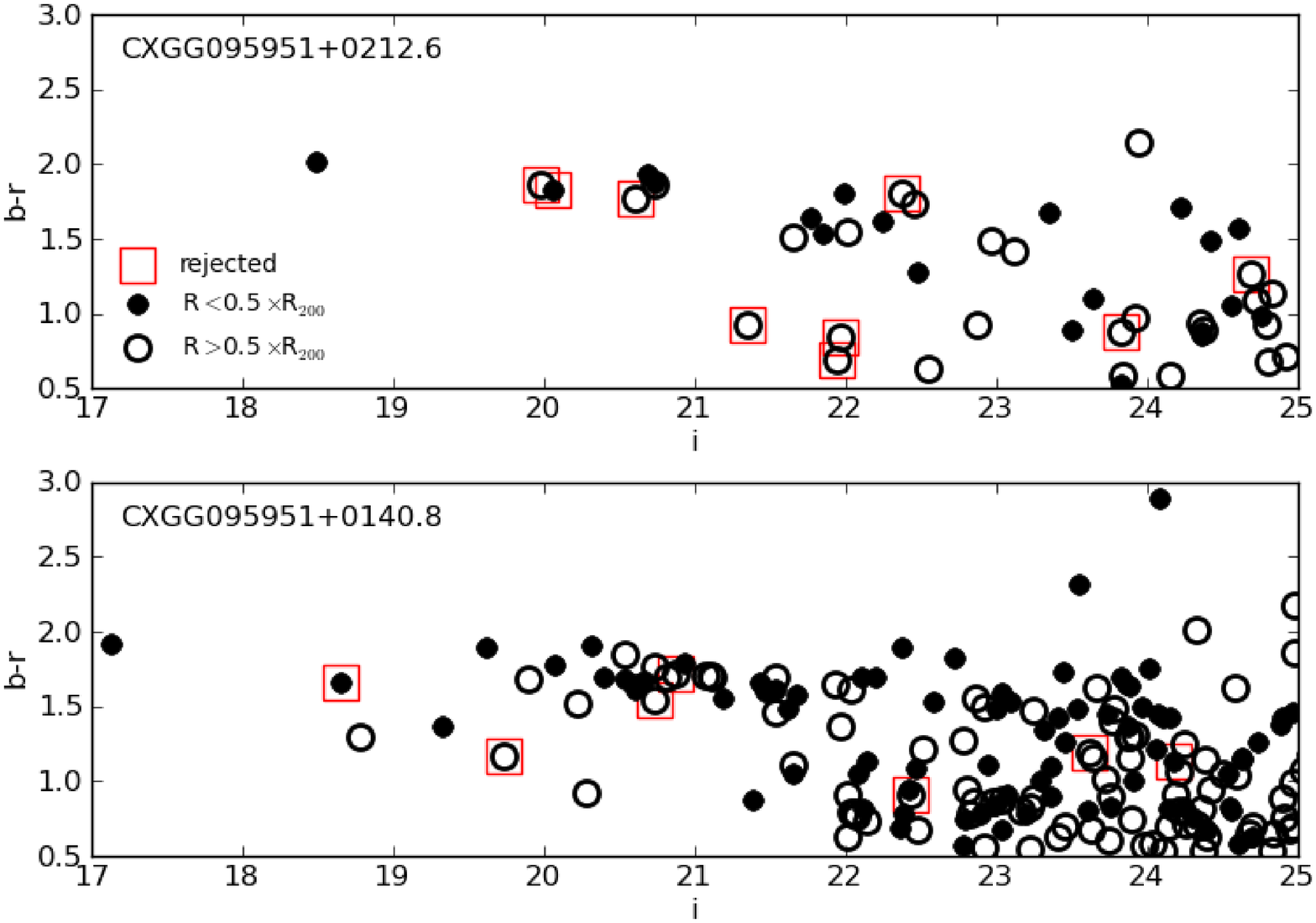}
\caption{$b - r$ vs $i$ colour--magnitude diagram (observed frame) for all member-candidate galaxies of either fossil group (black circles); galaxies within a group-centric distance of $0.5 R_{200}$ are marked by filled circles. Red open squares mark those photometric-member galaxies with a spectroscopic redshift inconsistent with that of either group (see Sect. 2).}
\end{figure*}

\subsection{Colour--magnitude diagram}

The dominance of passively evolving or quiescent galaxies
at $M^{\mathrm{stars}} \ge 10^{10}~\mathrm{M}_{\sun}$ is confirmed,
in a more direct way, by the $b - r$ vs $i$ colour--magnitude diagram
of either fossil group (see Fig. 4).
This diagram is obtained using Subaru photometry
(see Ilbert et al. 2009).
For $z \sim 0.4$, the $b - r$ colour nicely brackets
the redshifted spectral feature at $\sim 4000$~\AA~(rest frame),
whose amplitude increases together with the time since the last main episode
of star formation but also with increasing metallicity (e.g. Worthey 1994).

Both fossil groups exhibit a well defined red-sequence,
and a lack of galaxies within $0.5 R_{200}$ that are bluer
than this locus by $\ge 0.5~\mathrm{mag}$.
Most of these blue galaxies are at least $\sim 4$ magnitudes fainter
than the BCG of either fossil group in the $i$-band.
Moving out to $R_{200}$, the bona fide member galaxies with
a $b - r$ colour at least 0.5 mag bluer than the BCG one are still a few.
This confirms that most of the luminous galaxies
within the bound region of either fossil group are old and passively evolving
or quiescent, modulo dust attenuation (cf. Wilman et al. 2008).

The dominance of galaxies with a very low star-formation activity
among those that are about two to four magnitudes fainter than the BCG
is confirmed for the fossil groups RXC\,J2315.7$-$0222 and RXC\,J0216.7$-$4749
(D\'emocl\`es et al. 2010), AWM4 (Zibetti et al. 2009;
see fig.~2 in Koranyi \& Geller 2002), RX\,J1416.4$+$2315
(see fig.~2 in Cypriano, Mendes de Oliveira, \& Sodr\'e 2006),
and the fossil group candidate RX\,J1520.9$+$4840\footnote{For this object the virial radius was estimated on the basis of a scaling relation (Evrard, Metzler \& Navarro 1996).} (Eigenthaler \& Zeilinger 2009).
However, in the literature there is no comparative study
of the galaxy populations of fossil and non-fossil systems of similar masses
down to such fainter magnitudes, based on multi-band photometry
or spectroscopy.

Finally, we note that two member-candidate galaxies with $i \approx 23.4$
and $i \approx 24.1$ exhibit colors that are much redder than average color
of the red-sequence of CXGG\,095951$+$0140.8 ($b - r \approx 2.3$
and $b - r \approx 2.9$, respectively, vs $b - r \approx 1.7$).
These two very faint and red galaxies (about 6.3--$7~i$-mag fainter
and 0.4--$1~\mathrm{mag}$ redder in $b - r$ than the BCG) are located
at projected distances of 110 and $330~\mathrm{kpc}$ from the BCG
(or the cluster center), respectively.
Member-candidate galaxies with analogous photometric properties exist
in the fossil group candidate RX\,J1548.9$+$0851\footnote{The fossil nature of this group comes under scrutiny however (P. Eigenthaler private communication).} (Eigenthaler \& Zeilinger 2009).

\section[]{Discussion}

\subsection{Characteristic timescales}

In order to understand why the bulk of the galaxies
with $M^{\mathrm{stars}} \ge 10^{10}~\mathrm{M}_{\sun}$
are dominated by old, passively evolving stellar populations
in the two fossil groups at $z \approx 0.4$,
we need to estimate if there was enough time for the last formed stars
to age once their host galaxy had reached the bound region of either group.
Hence, we need to estimate the following timescales:
the time elapsed since half the DM halo of each fossil group
was in place (i.e., the so-called ``assembly time''),
and the time elapsed since the last significant infall of blue galaxies
with $M^{\mathrm{stars}} \ge 10^{10}~\mathrm{M}_{\sun}$ occurred.
These two timescales need to be compared
with the minimum life-time of the stellar populations
characterizing old, passively evolving (i.e., red-sequence) galaxies,
which is about 1--$2~\mathrm{Gyr}$.

We now make the conservative assumption that the two fossil groups
behave as other similarly massive groups observed at $z \approx 0.4$.
Furthermore, we assume that groups do not significantly increase
their total masses between $z \approx 0.5$ and $z \approx 0.4$\footnote{According to the adopted cosmological model (Sect.~1), this redshift range corresponds to a time interval of $0.7~\mathrm{Gyr}$.}, in a statistical sense.
By doing so, we can use results on the median formation redshifts
of DM haloes obtained from the simulations of Gao et al. (2004)
or different analytical models that are reproduced in fig.~5
of Giocoli et al. (2007).
Groups of similar total masses to CXGG\,095951$+$0212.6
and CXGG\,095951$+$0140.8 have assembled half of their DM halo
at typical redshifts of $\sim 1.2$ and $\sim 1.05$, respectively,
which correspond to similar lookback times of $\approx 4~\mathrm{Gyr}$
at $z = 0.425$ and $z = 0.372$.

According to the computations of dynamical friction and galaxy-merging
timescales of Boylan-Kolchin, Ma \& Quataert (2008, their fig. 1),
galaxies with $M^{\mathrm{stars}} \sim 10^{10}$
or $10^{11}~\mathrm{M}_{\sun}$ need about 6 or 3 Gyr, respectively,
to reach an orbit with maximum group-centric distance of $0.5 R_{200}$
from $R_{200}$ for the first time.
Since the simulations of Boylan-Kolchin et al. mostly neglect
the effect of the central galaxy in a given DM halo
on the merging time-scale, and do not take into account the drag
on ambient gas (e.g. Ostriker 1999), we infer that the epoch of last infall
of blue galaxies with $M^{\mathrm{stars}} \ge 10^{10}~\mathrm{M}_{\sun}$
can date back to less than 3--6 Gyr with respect to the lookback time
of either fossil group.

The time of last infall of blue, intermediate-mass galaxies
is comparable to the median time elapsed
since the assembly of a group as massive as either fossil group,
if not longer than that.
Thus, we can assume that most of the galaxies
with $M^{\mathrm{stars}} \ge 10^{10}~\mathrm{M}_{\sun}$
have resided within the bound region of either fossil group
since its assembly (cf. von Benda-Beckmann et al. 2008; Dariush et al. 2010).
This ``residence'' time of $\approx 4~\mathrm{Gyr}$ is longer
than the typical life-times of young stars, which are responsible
for the bulk luminosity shortward of the 4000~\AA-break,
when they are present.
Hence, there is enough time to observe
a dominating population of old, passively evolving galaxies
with $M^{\mathrm{stars}} \ge 10^{10}~\mathrm{M}_{\sun}$
in the two fossil groups at $z \approx 0.4$.
This is even more the case if the assembly time of a fossil group is shorter
than the median assembly time for all coeval, similarly massive groups
(D'Onghia et al. 2005; Dariush et al. 2007).
Environmental effects could be responsible for the raise of the population
of old, passively evolving galaxies in the two fossil groups
(see Boselli \& Gavazzi 2006 for a recent review).

\subsection{Star formation history}

Member-candidate galaxies of either fossil group
which are satellites at $z \approx 0.4$
could reach stellar masses of 1--$8 \times 10^{10}~\mathrm{M}_{\sun}$
before their star formation turned to its end.
We note that, in spite of the generally held, negative feedback
exerted by the group environment on the star formation activity
of member galaxies, such an activity seems to be enabled in 20\%--50\%
of the galaxies with $M^{\mathrm{stars}} \sim 10^{11}~\mathrm{M}_{\sun}$
in groups at similar redshifts and with similar masses (Fig.~2;
Giodini et al. in preparation; see also Wilman et al. 2008
and references therein).

It is easy to understand how tidal interactions\footnote{Tidal interaction may have been active in a de-localized way, in particular during the process of rapid and violent formation of the dominant galaxy (D'Onghia et al. 2005; see also Pierini et al. 2008 for an analogy), but, also, in a more localized way, since the early formation of the central dominant galaxy (see Elmegreen et al. 2000 for an analogy).} (e.g., Spitzer \& Baade 1951; Richstone 1976;
Farouki \& Shapiro 1981; Merritt 1983, 1984; Icke 1985; Miller 1986;
Byrd \& Valtonen 1990; Valluri \& Jog 1990) or harassment (Moore et al. 1996)
could have inhibited the growth of satellite galaxies
up to $M^{\mathrm{stars}} \sim 10^{10}~\mathrm{M}_{\sun}$.
It is less certain how the same interaction processes\footnote{Tidal interactions and harassment describe, respectively, prolonged and collective impulsive interactions of type galaxy--galaxy or galaxy--gravitational potential well of the parent system.} could have fostered the growth of more massive
satellite galaxies (e.g., Keel et al. 1985; Kennicutt et al. 1987;
Byrd \& Valtonen 1990; Mihos et al. 1992; Henriksen \& Byrd 1996;
Lake, Katz \& Moore 1998; Iono, Yun \& Mihos 2004;
Braglia, Pierini \& B\"ohringer 2007).

An earlier assembly time of fossil groups
with respect to non-fossil groups of similar masses
(D'Onghia et al. 2005; Dariush et al. 2007)
can justify the lower fraction of star-forming galaxies
with $10^{10} \le M^{\mathrm{stars}} \le 10^{11}~\mathrm{M}_{\sun}$
(see Sect.~4.1).
As discussed by D'Onghia et al. (2005), this is also consistent
with the larger X-ray-to-optical luminosity ratios of fossil groups
with respect to similarly massive non-fossil groups in the local Universe
claimed by Jones et al. (2003).
In fact, in groups with earlier assembly times
the hot gas was collected and started cooling earlier
and at higher central densities.
However, no difference in the scaling relation
between X-ray and optical luminosities is reported
for a sample of seven fossil groups at $z \le 0.2$
and total masses $M_{200}$ of a few times $10^{14}~\mathrm{M}_{\sun}$
(Voevodkin et al. 2010), at variance with previous results
(Jones et al. 2003; Khosroshahi et al. 2007).
As mentioned in Sect.~2, the X-ray-to-optical luminosity ratio
is rather high for CXGG\,095951$+$0140.8 but normal for CXGG\,095951$+$0212.6,
when compared to those of similarly massive groups at $z = 0.3$--0.5.
%Unfortunately, from the existing X-ray data, we can not measure
%the temperature of the hot intragroup medium of either fossil group,
%which would be a proxy for the depth of the potential well
%and, thus, the past dynamical history of the system.

Interestingly, there is growing evidence that
an environment characterized by high galaxy number density
and low galaxy velocity dispersion, such as the one found in compact groups,
plays a key role in accelerating galaxy evolution
by enhancing star-formation processes in galaxies
and favoring a fast transition to quiescence
(Tzanavaris et al. 2010; Walker et al. 2010 and references therein).
Such an environment could represent the immediate progenitor
of a fossil system.
This is straightforward if such a progenitor is a compact group,
and compact groups are interpreted as transient phases
in the dynamical evolution of looser groups (Barnes 1989).
Alternatively, an environment with the previous characteristics
can be produced if the existence of fossil systems is primarily driven
by the relatively early infall of massive satellites (D'Onghia et al. 2005;
von Benda-Beckmann et al. 2008).
Tentatively this may be achieved through preferential infall on orbits
with typically lower angular momentum.
Consistently, there is an indication of the presence of galaxies
in radial orbits in the external region of the fossil cluster
RX\,J105453.3$+$552102 (Aguerri et al. 2011).

In a future study, we will explore if an efficient confinement
of the hot, X-ray emitting gas (Mathews et al. 2005) could foster
the rather large stellar mass fractions of the two fossil groups under study
(cf. G09).

\subsection{Fossil group progenitors}

An earlier assembly time for fossil groups with respect to as massive
non-fossil groups is consistent with the observational results
discussed in Sect.~3; it is also predicted by simulations
in a statistical sense (D'Onghia et al. 2005; Dariush et al. 2007).
The evidence is a GSMF dominated by old, passively evolving systems
at $M^{\mathrm{stars}} \ge 10^{10}~\mathrm{M}_{\sun}$ (Sect.~3.1)
plus the corresponding existence of a tight red-sequence (Sect.~3.2).
Since the two fossil groups at $z \approx 0.4$ sit in very different
density environments of the LSS (Sect.~2), the conditions leading to
the formation of a fossil system, in particular to the shaping
of its galaxy population, do not appear to be intimately connected to
the surrounding LSS.
This is also consistent with simulations (von Benda-Beckmann et al. 2008;
Cui et al. 2011).

An early formation time was advocated by Aguerri et al. (2011)
to explain the dynamics and luminosity function of the member galaxies
as well as the photometric properties of the BCG
for the relaxed, massive ($M_{200} \sim 1 \times 10^{15}~\mathrm{M}_{\sun}$)
fossil system RX\,J105453.3$+$552102 at $z = 0.47$.
On the other hand, a deep optical image of the BCG
of the low X-ray luminosity fossil-group candidate RX\,J1548.9$+$0851
(Santos, Mendes de Oliveira \& Sodr\'e 2007) clearly reveals
the existence of many shells (Eigenthaler \& Zeilinger 2009).
According to Hernquist \& Spergel (1992), many shell systems
among elliptical and spheroidal galaxies may have originated
through major mergers of disc galaxies with comparable masses
rather than via minor mergers - between a massive elliptical galaxy
and a small disc companion (Quinn 1984; Hernquist \& Quinn 1988, 1989)
or a small elliptical/spiral companion (Dupraz \& Combes 1986) - or accretions
(Thomson \& Wright 1990; Thomson 1991).
We speculate that the relatively early infall of massive satellites
can better justify the formation of a rather massive fossil group
in a denser region of the LSS, like CXGG\,095951$+$0140.8,
whereas compact groups can more likely be the progenitors
of observed low-mass and isolated fossil groups, like CXGG\,095951$+$0212.6.

On one hand, the LSS surrounding the progenitor of a fossil group
could be responsible more for the growth of the total mass of the system
than for the growth of the stellar mass of its BCG
(see Khosroshahi, Ponman \& Jones 2006).
This is consistent with both the shallow dependence
of the median luminosity of BCGs on the total mass of the parent halo
observed in the local Universe (Yang, Mo \& van den Bosch 2008)
and the result from the hydrodynamic simulations of Cui et al. (2011)
that X-ray bright groups are disproportionally rare in low-density regions.

On the other hand, it is reasonable to assume that isolated fossil groups
can be identified as such for longer times than non isolated ones
of similar mass, because it is more difficult
either that their magnitude gap is replenished by infalling galaxies,
or that they fall into larger bound structures.
In a low density environment, a loose group has more chances
to keep its identity and dynamically evolve into a fossil group
after a compact group phase (cf. Barnes 1989).
Either in the case of early assembly (D'Onghia et al. 2005;
Dariush et al. 2007) or in the case of final evolution of a loose group
(Barnes 1989), a rapid build-up of the stellar mass
and the following quick end of the star-formation activity
in most of the brightest member galaxies need to be invoked.
Interestingly, this is what compact groups in the local Universe
seem to have experienced (Tzanavaris et al. 2010; Walker et al. 2010
and references therein).

If two channels lead to the formation of a fossil group
and there is a sort of density environment bias
on identification and inference on the progenitor from observations,
the stellar populations of the BCG of a fossil group
do not necessarily have to be remarkably different
from those of field, massive early-type galaxies of similar optical
and X-ray luminosities at $z \sim 0$ (cf. La Barbera et al. 2009).
Furthermore, the size of environmental effects on typical red-sequence galaxies
is still a controversial issue, and might be quite small
at large stellar masses (cf. e.g., Thomas et al. 2005, 2010;
Pannella et al. 2009; Cooper et al. 2010).
Hence, there might be less tension between our conclusions
and the result of La Barbera et al. (2009)\footnote{In addition, we note that comparison of models and observations is not straightforward even when considering simple stellar population models (Conroy \& Gunn 2010 and references therein).}.

Future multi-wavelength studies of statistical samples of fossil groups
with different masses and redshifts (e.g., Santos et al. 2007),
located in different density regions of the LSS,
will provide a test of our interpretation and hypotheses.

\subsection{Intracluster light}

Here we discuss if the contribution from low-surface brightness, diffuse
stellar emission not associated with galaxies (the so-called intracluster
light - ICL) could explain the rather low value of $L_i/L_{{\mathrm X}}$
found in Sect.~2 for the more massive fossil group, CXGG\,095951$+$0140.8.
The impact on the total stellar mass budget of the ICL
has already been pointed out in the literature, and with different scales,
from observations (cf. e.g., Zibetti et al. 2005 and Gonzalez et al. 2007).
In simulations, most of the ICL seems to be associated with the build-up
of the BCG (e.g., Murante et al. 2007 and references therein).
The expected amount of ICL is uncertain as well as the observed one - 3\%
to 30\% of all stars in groups at $z = 0$ (e.g., Kapferer et al. 2010
and references therein) - and seems to be independent of the mass-scale
of the galaxy system (e.g., Dolag, Murante \& Borgani 2010
and references therein).

Groups of galaxies that formed early (as fossil groups, potentially)
might be particularly effective at producing ICL,
as interactions in the group environment begin to liberate tidal material.
Galaxy systems as massive as the two fossil groups under study
have smaller velocity dispersions than do massive clusters,
meaning that interactions on the group scales can be slow and damaging
as opposite to the impulsive high speed encounters,
e.g. by harrassment, on the cluster scale.
These slow tidal interactions can strip material
from rotating disc galaxies into loosely bound tidal structures,
forming long tails of stars and gas (D'Onghia et al. 2009).
It is then interesting to estimate this additional amount of unbound stars
that can escape detection in the present data.

In particular, we use the approximations developed in D'Onghia et al. (2010)
to compute the fraction of ICL that a disc galaxy with a dynamical mass
equal to $5 \times 10^{11}~\mathrm{M}_{\sun}$
(of which $2 \times 10^{10}~\mathrm{M}_{\sun}$ is in the disc) produces
passing on a parabolic orbit within the potential of a typical group.
Such a galaxy has a stellar mass of the order of $10^{10}~\mathrm{M}_{\sun}$.
Figure 5 shows that the smaller the pericentric distance
the larger the fraction of light stripped into the intergalactic medium.
The fraction of ICL computed for the model galaxy
is given by the ratio of the number of unbound stars
to the total amount of stars bound to all the galaxies of the group.
This galaxy, passing at 30-40 kpc from the group center,
can contribute up to 3\% of the total light in the group galaxies.
Hence, slow tidal interactions likely provide a marginal contribution
to the total stellar budget of a low-mass fossil group
as CXGG\,095951$+$0212.6, but a non-neglible one for a fossil group
as massive as CXGG\,095951$+$0140.8.
Given these rough estimates, we will estimate the ICL contribution
to the total stellar luminosity of the two fossil groups under study
in the future, thanks to the deep B- and R-band photometry
recently obtained by us with the Wide Field Imager (Baade et al. 1999),
mounted at the Cassegrain focus of the MPG/ESO $2.2~\mathrm{m}$ telescope
in La Silla, Chile.

\begin{figure}
\includegraphics[width=84mm]{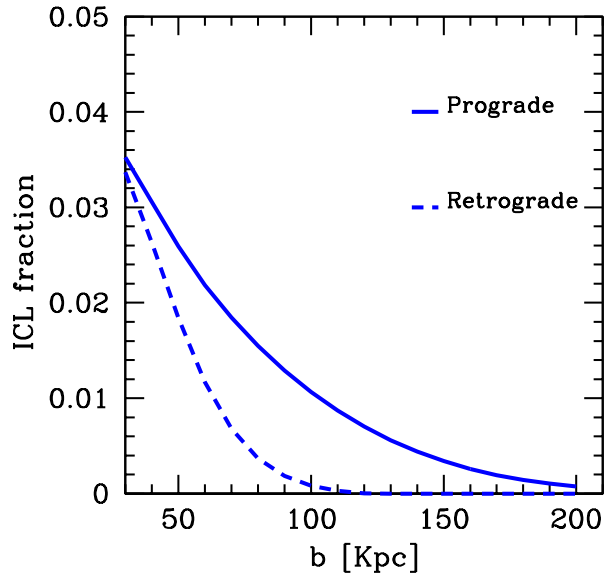}
\caption{The fraction of ICL released by the model galaxy (see text) orbiting within the potential of a typical group computed as a function of pericentric distance (from D'Onghia et al. 2010). Due to the quasi-resonance phenomenon of the tidal interaction, the prograde encounters are more efficient in stripping material than the retrograde ones.}
\end{figure}

\section[]{Conclusions}

For the first time, the galaxy stellar mass function (GSMF) was computed,
down to a stellar mass $M^{\mathrm{stars}} \sim 10^{9}~\mathrm{M}_{\sun}$
and in a robust statistical way, for two fossil groups
discovered at $z \approx 0.4$: CXGG\,095951$+$0212.6 and CXGG\,095951$+$0140.8.
These groups have also well-determined total masses ($M_{200}$)
equal to $1.9~(\pm 0.41) \times 10^{13}$
and $9.5~(\pm 0.42) \times 10^{13}~\mathrm{M}_{\sun}$, respectively.
This was possible because of the high-quality multi-wavelength information
available for the $2~\mathrm{deg}^2$ area of the Cosmic Evolution Survey
({\it COSMOS\/}; Scoville et al. 2007) where the two systems
were initially detected in X-rays (Finoguenov et al. in preparation; G09).
In particular, the identification as fossil
and the selection of the member-candidate galaxies of either group
rest on robust photometric redshifts,
and $i_{\mathrm AB} \le 22.5~\mathrm{mag}$ limited spectroscopy.

Interestingly, the large-scale structure (LSS)
in which the two fossil groups are embedded is very different:
CXGG\,095951$+$0212.6 appears isolated,
whereas CXGG\,095951$+$0140.8 belongs to a LSS
that covers the entire $2~\mathrm{deg}^2$ area of the {\it COSMOS\/}
(corresponding to a cross size of about $25.5~\mathrm{Mpc}$ at $z \approx 0.4$)
and is traced by a total of 28 X-ray emitting groups.

Comparison of the GSMF of these fossil groups
and the composite GSMF of the {\it COSMOS\/} X-ray emitting groups
at $0.3 \le z \le 0.5$ that span a similar range in total mass
(Giodini et al. in preparation) highlights a lack of star-forming galaxies
with $M^{\mathrm{stars}} \ge 10^{10}~\mathrm{M}_{\sun}$
in the two fossil groups.
Consistently, the distribution of their photometric member galaxies
in the $b - r$ vs $i$ colour--magnitude diagram
reveals a well-defined red sequence and a lack of significantly bluer,
luminous galaxies out to the virial radius $R_{200}$
of either fossil group.
In addition, the total stellar mass fraction of these groups
within $0.7 R_{200}$ is relatively large for their total masses (G09),
in particular for the less massive one (i.e., CXGG\,095951$+$0212.6).

This evidence suggests that the overall conversion of (cold) gas
into stars was accelerated and/or more efficient in fossil groups
with respect to the average X-ray emitting group
in the same mass range and at similar redshift.
The star-formation activity in member galaxies rapidly ceased afterwards.
At the same time, no significant infall of galaxies
with $M^{stars} \ge 10^{10}~\mathrm{M}_{\sun}$
took place in the last 3--6 Gyr for either fossil group at $z \approx 0.4$.

If the processes shaping the galaxy component of a fossil group
seem to be independent of the LSS, the progenitor of an observed fossil group
may depend on it.
On the basis of the measured stellar mass fractions (G09)
and X-ray-to-optical luminosity ratios (Giodini et al. in preparation),
we propose that compact groups are the most likely progenitors
of observed low mass and isolated fossil groups, like CXGG\,095951$+$0212.6.
Conversely, the relatively early infall of massive satellites likely originates
massive fossil groups in dense regions of the LSS, like CXGG\,095951$+$0140.8.

\section*{Acknowledgments}

%I thank Professor N. Kameswara Rao for some helpful suggestions,
%Dr H. C. Bhatt for a critical reading of the original version of the
%paper and an anonymous referee for very useful comments that improved
%the presentation of the paper.
We thank the referee, Paul Eigenthaler, for a careful reading
of the original version of the paper and the useful comments
that improved the presentation of the paper.
DP acknowledges the kind and fruitful hospitality
at the Max-Planck-Institut f\"ur extraterrestrische Physik (MPE).

\bsp

\label{lastpage}


\begin{thebibliography}{99}
\bibitem[\protect\citeauthoryear{Aguerri et al.}{2011}]{aguerri11} Aguerri J.A.L. et al. 2011, A\&A, 527, 143
\bibitem[\protect\citeauthoryear{Arimoto \& Yoshii}{}]{arimoto87} Arimoto N., Yoshii, Y. 1987, A\&A, 173, 23
\bibitem[\protect\citeauthoryear{Baade et al.}{1999}]{baade99} Baade D. et al. 1999, Msgnr, 95, 15
\bibitem[\protect\citeauthoryear{Balogh et al.}{2000}]{balogh00} Balogh M.L., Navarro J.F., Morris S.L., 2000, ApJ, 540, 113
\bibitem[\protect\citeauthoryear{Barnes}{1989}]{barnes89} Barnes J.E., 1989, Nature, 338, 123
\bibitem[\protect\citeauthoryear{Boselli \& Gavazzi}{2006}]{boselli06} Boselli A., Gavazzi G., 2006, PASP, 118, 517
\bibitem[\protect\citeauthoryear{Boylan-Kolchin et al.}{2008}]{boylan-kolchin08} Boylan-Kolchin M., Ma C.-P., Quataert E., 2008, MNRAS, 383, 93
\bibitem[\protect\citeauthoryear{Braglia et al.}{2007}]{braglia07} Braglia F.G., Pierini D., B\"ohringer H., 2007, A\&A, 470, 425
\bibitem[\protect\citeauthoryear{Braglia et al.}{2009}]{braglia09} Braglia F.G., Pierini D., Biviano A., B\"ohringer H., 2009, A\&A, 500, 947
\bibitem[\protect\citeauthoryear{Byrd \& Valtonen}{1990}]{byrd90} Byrd G., Valtonen M., 1990, ApJ, 350, 89
\bibitem[\protect\citeauthoryear{Chabrier}{2003}]{chabrier03} Chabrier G., 2003, ApJ, 586, L133
\bibitem[\protect\citeauthoryear{Conroy \& Gunn}{2010}]{conroy10} Conroy C., Gunn J.E., 2010, ApJ, 712, 833
\bibitem[\protect\citeauthoryear{Cooper et al.}{2010}]{cooper10} Cooper M.C., Gallazzi A., Newman J.A., Yan R., 2010, MNRAS, 402, 1942
\bibitem[\protect\citeauthoryear{Cui et al.}{2011}]{cui11} Cui W., Springel V., Yang X., De Lucia G., Borgani S, 2011, MNRAS, in press (arXiv:1102.4269)
\bibitem[\protect\citeauthoryear{Cypriano et al.}{2006}]{cypriano06} Cypriano E.S., Mendes de Oliveira C.L., Sodr\'e L. Jr., 2006, AJ, 132, 514
\bibitem[\protect\citeauthoryear{Dariush et al.}{2007}]{dariush07} Dariush A., Khosroshahi H.G., Ponman T.J., Pearce F., Raychaudhury S., Hartley W., 2007, MNRAS, 382, 433
\bibitem[\protect\citeauthoryear{Dariush et al.}{2010}]{dariush10} Dariush A., Raychaudhury S., Ponman T.J., Khosroshahi H.G., Benson A.J., Bower R.G., Pearce F., 2010, MNRAS, 405, 1873
\bibitem[\protect\citeauthoryear{Democles et al.}{2010}]{democles10} D\'emocl\`es J., Pratt G.W., Pierini D., Arnaud M., Zibetti S., D'Onghia E., 2010, A\&A, 517, 52
\bibitem[\protect\citeauthoryear{Dolag et al.}{2010}]{dolag2010} Dolag K., Murante G., \& Borgani S., 2010, MNRAS, 405, 1544
\bibitem[\protect\citeauthoryear{D'Onghia \& Lake}{2004}]{donghia04} D'Onghia E., Lake G., 2004, ApJ, 612, 628
\bibitem[\protect\citeauthoryear{D'Onghia et al.}{2005}]{donghia05} D'Onghia E., Sommer-Larsen J., Romeo A.D., Burkert A., Pedersen K., Portinari L., Rasmussen J., 2005, ApJ, 630, L109
\bibitem[\protect\citeauthoryear{D'Onghia et al.}{2009}]{donghia09} D'Onghia E., Besla G., Cox T.J., Hernquist L., 2009, Nature, 460, 605
\bibitem[\protect\citeauthoryear{D'Onghia et al.}{2010}]{donghia10} D'Onghia E., Vogelsberger M., Figuere-Ciguere C., Hernquist L., 2010, ApJ, 725, 353
\bibitem[\protect\citeauthoryear{Dunkley et al.}{2009}]{dunkley09} Dunkley J. et al., 2009, ApJS, 180, 306
\bibitem[\protect\citeauthoryear{Dupraz \& Combes}{1986}]{dupraz86} Dupraz C., Combes F., 1986, A\&A, 166, 53
\bibitem[\protect\citeauthoryear{Eigenthaler \& Zeilinger}{2009}]{eigenthaler09} Eigenthaler P., Zeilinger W. W., 2009, Astronomische Nachrichten, 330, 978
\bibitem[\protect\citeauthoryear{Ellingson et al.}{2001}]{ellingson01} Ellingson E., Lin H., Yee H.K.C., Carlberg R.G., 2001, ApJ, 547, 609
\bibitem[\protect\citeauthoryear{Elmegreen et al.}{2000}]{elmegreen00} Elmegreen D.M., Elmegreen B.G., Chromey F.R., Fine M.S., 2000, AJ, 120, 733
\bibitem[\protect\citeauthoryear{Evrard et al.}{1996}]{evrard96} Evrard A.E., Metzler C.A., Navarro J.F., 1996, ApJ, 469, 494
\bibitem[\protect\citeauthoryear{Farouki \& Shapiro}{1981}]{farouki81} Farouki R., Shapiro S.L., 1981, ApJ, 243, 32
\bibitem[\protect\citeauthoryear{Finoguenov et al.}{2007}]{finoguenov07} Finoguenov A. et al. 2007, ApJS, 172, 182
\bibitem[\protect\citeauthoryear{Gao et al.}{2004}]{gao04} Gao L., White S.D.M., Jenkins A., Stoehr F., Springel V., 2004, MNRAS, 355, 819
\bibitem[\protect\citeauthoryear{Giocoli et al.}{2007}]{giocoli07}
Giocoli C., Moreno J., Sheth R.K., Tormen, G., 2007, MNRAS, 376, 977
\bibitem[\protect\citeauthoryear{Giodini et al.}{2009}]{giodini09} Giodini S. et al. 2009, ApJ, 703, 982 (G09)
\bibitem[\protect\citeauthoryear{Giodini et al.}{2010}]{giodini10} Giodini S. et al. 2010, ApJ, 714, 218
\bibitem[\protect\citeauthoryear{Gonzalez et al.}{2007}]{gonzalez07} Gonzalez A.H., Zabludoff A.I., Zaritsky D., 2007, ApJ, 666, 147
\bibitem[\protect\citeauthoryear{Henriksen \& Byrd}{1996}]{henriksen96} Henriksen M., Byrd, G., 1996, ApJ, 459, 82
\bibitem[\protect\citeauthoryear{Hernquist \& Quinn}{1988}]{hernquist88} Hernquist L., Quinn P.J., 1988, ApJ, 331, 682
\bibitem[\protect\citeauthoryear{Hernquist \& Quinn}{1989}]{hernquist89} Hernquist L., Quinn P.J., 1989, ApJ, 342, 1
\bibitem[\protect\citeauthoryear{Hernquist \& Spergel}{1992}]{hernquist92} Hernquist L., Spergel D.N., 1992, ApJ, 399, L117
\bibitem[\protect\citeauthoryear{Icke}{1985}]{icke85} Icke V., 1985, A\&A, 144, 115
\bibitem[\protect\citeauthoryear{Ilbert et al.}{2009}]{ilbert09} Ilbert O. et al., 2009, ApJ, 690, 1236
\bibitem[\protect\citeauthoryear{Iono et al.}{2004}]{iono04} Iono D., Yun M.S., Mihos J.C., 2004, ApJ, 616, 199
\bibitem[\protect\citeauthoryear{Jones et al.}{2000}]{jones00} Jones L.R., Ponman T.J., Forbes D.A., 2000, MNRAS, 312, 139
\bibitem[\protect\citeauthoryear{Jones et al.}{2003}]{jones03} Jones L.R., Ponman T.J., Horton A., Babul A., Ebeling H., Burke D.J., 2003, MNRAS, 343, 627
\bibitem[\protect\citeauthoryear{Kapferer et al.}{2010}]{kapferer10} Kapferer W., Schindler S., Knollmann S.R., van Kampen E., 2010, A\&A, 516, 41
\bibitem[\protect\citeauthoryear{Keel et al.}{1985}]{keel85} Keel W.C., Kennicutt R.C. Jr., Hummel E., van der Hulst J.M., 1985, AJ, 90, 708
\bibitem[\protect\citeauthoryear{Kennicutt et al.}{1987}]{kennicutt87} Kennicutt R.C. Jr., Roettiger K.A., Keel W.C., van der Hulst J.M., Hummel E., 1987, AJ, 93, 1011
\bibitem[\protect\citeauthoryear{Khosroshahi et al.}{2006}]{khosroshahi06} Khosroshahi H.G., Ponman, T.J., Jones L.R., 2006, MNRAS, 372, 68
\bibitem[\protect\citeauthoryear{Khosroshahi et al.}{2007}]{khosroshahi07} Khosroshahi H.G., Ponman, T.J., Jones L.R., 2007, MNRAS, 377, 595
\bibitem[\protect\citeauthoryear{Knobel et al.}{2009}]{knobel07} Knobel C. et al., 2009, ApJ, 697, 1842
\bibitem[\protect\citeauthoryear{Koekemoer et al.}{2007}]{koekemoer07} Koekemoer A.M. et al., 2007, ApJS, 172, 196
\bibitem[\protect\citeauthoryear{Komatsu et al.}{2009}]{komatsu09} Komatsu E. et al., 2009, ApJS, 180, 330
\bibitem[\protect\citeauthoryear{Koranyi \& Geller}{2002}]{koranyi02} Koranyi D.M., Geller M.J., 2002, AJ, 123, 100
\bibitem[\protect\citeauthoryear{La Barbera et al.}{2009}]{labarbera09} La Barbera F., de Carvalho R.R., de la Rosa I.G., Sorrentino G., Gal R.R., Kohl-Moreira J.L., 2009, AJ, 137, 3942
\bibitem[\protect\citeauthoryear{Lake et al.}{1998}]{lake1998} Lake G., Katz N., Moore B., 1998, ApJ, 495, 152
\bibitem[\protect\citeauthoryear{Leauthaud et al.}{2010}]{leauthaud10} Leauthaud A. et al., 2010, ApJ, 709, 97
\bibitem[\protect\citeauthoryear{Lilly et al.}{2007}]{lilly07} Lilly S.J. et al., 2007, ApJS, 172, 70
\bibitem[\protect\citeauthoryear{Mathews et al.}{2005}]{mathews05} Mathews W.G., Faltenbacher A., Brighenti F., Buote D.A., 2005, ApJ, 634, L137
\bibitem[\protect\citeauthoryear{Merritt}{1983}]{merritt83} Merritt D., 1983, ApJ, 264, 24
\bibitem[\protect\citeauthoryear{Merritt}{1984}]{merritt84} Merritt D., 1984, ApJ, 276, 26
\bibitem[\protect\citeauthoryear{Mihos}{1992}]{mihos92} Mihos J.C., Richstone D.O., Bothun G.D., 1992, ApJ, 400, 153
\bibitem[\protect\citeauthoryear{Miller}{1986}]{miller86} Miller R.H., 1986, A\&A, 167, 41
\bibitem[\protect\citeauthoryear{Milosavljevi\'c et al.}{2006}]{milosavljevic06} Milosavljevi\'c M., Miller C.J., Furlanetto S.R., Cooray A., 2006, ApJ, 637, L9
\bibitem[\protect\citeauthoryear{Moore et al.}{1996}]{moore96} Moore B., Katz N., Lake G., Dressler A., Oemler A., 1996, Nature, 379, 613
\bibitem[\protect\citeauthoryear{Murante et al.}{2007}]{murante07} Murante G., Giovalli M., Gerhard O., Arnaboldi M., Borgani S., Dolag K., 2007, MNRAS, 377, 2
\bibitem[\protect\citeauthoryear{Ostriker}{1999}]{ostriker99} Ostriker E.C., 1999, ApJ, 513, 252
\bibitem[\protect\citeauthoryear{Pannella et al.}{2009}]{pannella09} Pannella M. et al., 2009, ApJ, 701, 787
\bibitem[\protect\citeauthoryear{Pierini et al.}{2004}]{pierini04} Pierini D., Maraston C., Bender R., Witt A.N., 2004, MNRAS, 347, 1
\bibitem[\protect\citeauthoryear{Pierini et al.}{2008}]{pierini08} Pierini D., Zibetti S. Braglia F., B\"ohringer H., Finoguenov A., Lynam P.D., Zhang Y.-Y., 2008, A\&A, 483, 727
\bibitem[\protect\citeauthoryear{Ponman \& Bertram}{1993}]{ponman93} Ponman T.J., Bertram D., 1993, Nature, 363, 51
\bibitem[\protect\citeauthoryear{Ponman et al.}{1994}]{ponman94} Ponman T.J., Allan D.J., Jones L.R., Merrifield M., McHardy I.M., Lehto H.J., Luppino G.A., 1994, Nature, 369, 462
\bibitem[\protect\citeauthoryear{Quinn}{1984}]{quinn84} Quinn P.J., 1984, ApJ, 279, 596
\bibitem[\protect\citeauthoryear{Richstone}{1976}]{richstone76} Richstone D.O., 1976, ApJ, 204, 642
\bibitem[\protect\citeauthoryear{Romer et al.}{2000}]{romer00} Romer A.K. et al. 2000, ApJS, 126, 209
\bibitem[\protect\citeauthoryear{Sales et al.}{2007}]{sales07} Sales L.V., Navarro J.F. Lambas D.G., White S.D.M., Croton D.J., 2007, MNRAS, 382, 1901
\bibitem[\protect\citeauthoryear{Salpeter}{1955}]{salpeter55} Salpeter E.E., 1955, ApJ, 121, 161
\bibitem[\protect\citeauthoryear{Santos}{2007}]{santos07} Santos W.A., Mendes de Oliveira C., Sodr\'e L. Jr., 2007, AJ, 134, 1551
\bibitem[\protect\citeauthoryear{Schechter}{1976}]{schechter76} Schechter P., 1976, ApJ, 203, 297
\bibitem[\protect\citeauthoryear{Schirmer et al.}{2010}]{schirmer10} Schirmer M., Suyu S., Schrabback T., Hildebrandt H., Erben T., Halkola A., 2010, A\&A, 514, 60
\bibitem[\protect\citeauthoryear{Scoville et al.}{2007}]{scoville07} Scoville N. et al., 2007, ApJS, 172, 1
\bibitem[\protect\citeauthoryear{Spitzer \& Baade}{1951}]{spitzer51} Spitzer L. Jr., Baade W., 1951, ApJ, 113, 413
\bibitem[\protect\citeauthoryear{Sun et al.}{2009}]{sun09} Sun M., Voit G.M., Donahue M., Jones C., Forman W., Vikhlinin A., 2009, ApJ, 693, 1142
\bibitem[\protect\citeauthoryear{Teyssier et al.}{2011}]{teyssier11} Teyssier R., Moore B., Martizzi D., Dubois Y., Mayer L., 2011, MNRAS, 414, 195
\bibitem[\protect\citeauthoryear{Thomas et al.}{2005}]{thomas05} Thomas D., Maraston C., Bender R., Mendes de Oliveira C., 2005, ApJ, 621, 673
\bibitem[\protect\citeauthoryear{Thomas et al.}{2010}]{thomas10} Thomas D., Maraston C., Schawinsk, K., Sarzi M., Silk J., 2010, MNRAS, 404, 1775
\bibitem[\protect\citeauthoryear{Thomson}{1991}]{thomson91} Thomson R.C., 1991, MNRAS, 253, 256
\bibitem[\protect\citeauthoryear{Thomson \& Wright}{1990}]{thomson90} Thomson R.C., Wright A.E., 1991, MNRAS, 247, 122
\bibitem[\protect\citeauthoryear{Tully \& Shaya}{1984}]{tully84} Tully R.B., Shaya E.J., 1984, ApJ, 281, 31
\bibitem[\protect\citeauthoryear{Tzanavaris et al.}{2010}]{tzanavaris10} Tzanavaris P. et al., 2010, ApJ, 716, 556
\bibitem[\protect\citeauthoryear{Valluri \& Jog}{1990}]{valluri90} Valluri M., Jog C.J., 1990, ApJ, 357, 367
\bibitem[\protect\citeauthoryear{Vikhlinin et al.}{1999}]{vikhlinin99} Vikhlinin A., McNamara B.R., Hornstrup A., Quintana H., Forman W., Jones C., Way M., 1999, ApJ, 520, L1
\bibitem[\protect\citeauthoryear{Voevodkin et al.}{2010}]{voevodkin10} Voevodkin A., Borozdin K., Heitmann K., Habib S., Vikhlinin A., Mescheryakov A., Hornstrup A., Burenin R., 2010, ApJ, 708, 1376
\bibitem[\protect\citeauthoryear{von Benda-Beckmann et al.}{2008}]{vonbendabeckmann08} von Benda-Beckmann A.M., D'Onghia E., Gottl\"ober S., Hoeft M., Khalatyan A., Klypin A., M\"uller V., 2008, MNRAS, 386, 2345
\bibitem[\protect\citeauthoryear{Walker et al.}{2010}]{walker10} Walker L.M., Johnson K.E., Gallagher S.C., Hibbard J.E., Hornschemeier A.E., Tzanavaris P., Charlton J.C., Jarrett T.H., 2010, AJ, 140, 1254
\bibitem[\protect\citeauthoryear{Wilman et al.}{2008}]{wilman08} Wilman D.J. et al., 2008, ApJ, 680, 1009
\bibitem[\protect\citeauthoryear{Worthey}{1994}]{worthey94} Worthey G., 1994, ApJS, 95, 107
\bibitem[\protect\citeauthoryear{Yang et al.}{2008}]{yang08} Yang X., Mo H.J., van den Bosch F.C., 2008, ApJ, 676, 248
\bibitem[\protect\citeauthoryear{Zapata et al.}{2009}]{zapata09} Zapata T., Perez J., Padilla N., Tissera P., 2009, MNRAS, 394, 2229
\bibitem[\protect\citeauthoryear{Zibetti et al.}{2005}]{zibetti05} Zibetti S., White S.D.M., Schneider D.P., Brinkmann J. 2005, MNRAS, 358, 949
\bibitem[\protect\citeauthoryear{Zibetti et al.}{2009}]{zibetti09} Zibetti S., Pierini D., Pratt G.W. 2009, MNRAS, 392, 525
%\bibitem[\protect\citeauthoryear{}{}]{}
%\bibitem[\protect\citeauthoryear{}{}]{}
%\bibitem[\protect\citeauthoryear{}{}]{}
%\bibitem[\protect\citeauthoryear{}{}]{}
%\bibitem[\protect\citeauthoryear{}{}]{}
%\bibitem[\protect\citeauthoryear{}{}]{}
%\bibitem[\protect\citeauthoryear{}{}]{}
%\bibitem[\protect\citeauthoryear{}{}]{}
%\bibitem[\protect\citeauthoryear{}{}]{}
\end{thebibliography}
\end{document}